\documentclass[12pt]{iopart}
\usepackage{iopams}
\usepackage{graphicx}
\usepackage{color}
\begin{document}
\title[Exact microcanonical statistical analysis of transition behavior ...]%
{Exact microcanonical statistical analysis of transition behavior in Ising 
chains and strips}
\author{K Sitarachu$^1$, R K P Zia$^{2, 3, 4}$, and M Bachmann$^1$}
\address{$^1$ Soft Matter Systems Research Group, Center for Simulational 
Physics, Department of Physics and Astronomy, University of Georgia, Athens, 
GA 30602, USA}
\address{$^2$ Center for Soft Matter and Biological Physics, Department of Physics, Virginia Tech, Blacksburg, VA 24061, USA}
\address{$^3$ Department of Physics \& Astronomy, University of North Carolina at Asheville, Asheville, NC 28804, USA}
\address{$^4$ Physics Department, University of Houston, Houston, Texas 77204, USA}
\begin{abstract}
Recent analyses of least-sensitive inflection points in derivatives of the 
microcanonical entropy for the two-dimensional Ising model revealed 
higher-order transition signals in addition to the well-studied second-order 
ferromagnetic/paramagnetic phase transition. In this paper, we re-analyze the 
exact density of states for the one-dimensional Ising chain as well as 
the strips with widths/lengths of up to 64/1024 spins, in 
search of potential transition features. While some transitions begin 
to emerge as the strip width increases, none are found for the chain,
as might be expected.
\end{abstract}
\submitto{\JSTAT}
\maketitle
\section{Introduction}
The (Lenz-)Ising model~\cite{lenz1,ising1} is one of the simplest spin 
models for the study of ferromagnetic order in crystals. Conventional 
statistical analysis of the exactly solvable one-dimensional (1D)
model~\cite{ising1} did not reveal thermodynamic phase transition features at 
finite temperatures, though, despite significant energy fluctuations. However, 
the two-dimensional (2D) problem, which was first solved exactly by Onsager 
by means of canonical statistical analysis~\cite{onsager1}, exhibits 
common signatures of a second-order phase transition between the disordered 
paramagnetic and the ordered ferromagnetic phase. The simplicity of 
the model made it a popular standard model for investigating complexity and 
triggered a vast number of theoretical studies aimed at a more 
general understanding of thermodynamic phase transitions. It also 
significantly contributed to the development of computational statistical 
physics as advanced algorithmic methodologies such as Monte Carlo sampling 
and finite-size scaling strategies enabled a better understanding of how 
short-range interaction can cause long-range order under thermal conditions.

The canonical statistical analysis method used in most of these studies works 
particularly well for very large systems, where the assumption of negligible 
surface contact of the system of interest with the surrounding heat bath is 
justified and the heat bath temperature can be considered a suitable and 
adjustable parameter controlling the equilibrium properties of the system. 
One of the important consequences is that different response quantities and 
an appropriate order parameter indicate the catastrophic fluctuations 
accompanying a phase transition at a unique transition point. For finite 
systems, this is not the case and the identification of a transition 
\emph{point} is not generally possible; different response quantities 
suggest different transition points. It is therefore common to extrapolate 
data 
obtained in computer simulations, where only systems of finite size can be 
simulated, toward the thermodynamic limit by finite-size scaling. This is 
particularly useful for the analysis of second-order transitions where the 
system loses its identity close to the phase transition point in the 
thermodynamic limit. All fluctuating quantities are non-analytic at the 
same transition point and can be quantified by power laws with specific 
critical exponents. 

However, 
the recent enormous advances in the development of technologies that makes 
possible 
experiments and applications on nanoscales has increased the interest of 
interdisciplinary sciences like biochemistry in 
using statistical physics methods for the understanding of complex behavior 
and long-range cooperativity in \emph{finite systems}. Thus, 
the foundation of the statistical methods that have been so successful in 
understanding phase transitions in large systems needs to be extended to 
include systems such as proteins for which the thermodynamic limit is 
an inappropriate assumption~\cite{mb1}.

The recently introduced generalized microcanonical inflection-point 
method~\cite{qb1} was developed to overcome these issues. It ultimately 
enables the systematic identification and even classification of transition 
signals in systems of arbitrary size.

In this paper, we revisit the one-dimensional (1D) Ising spin chain and 
extend our study to Ising strips with $L \times M$ spins attached to the 
nodes of a square 
lattice with $L$ being the length and $M$ the finite width of the 
strips. It has long been known 
that the 1D Ising chain does not experience a thermodynamic phase 
transition at finite temperatures~\cite{ising1}. However, the 
specific heat curve exhibits 
interesting monotonic features. It is important to note that such features 
like peaks in quantities such as the specific heat
are indeed often used as signals of cooperative behavior in systems of finite 
size, in particular those that do not possess a thermodynamic limit 
(e.g., finitely long, heterogeneous systems such as proteins). Since the 
thermodynamic 
limit represents an artificial situation, canonical statistical analysis  
faces the dilemma of not being able to rigorously distinguish transitions 
signatures in finite systems that might correspond to phase 
transitions in the thermodynamic limit from those that do not. For this 
reason, it is instructive to consider the exact, microcanonical solution
of the 1D Ising chain first. We then study the properties of the more 
interesting 2D systems, i.e., the periodic Ising strips of length 
$L$ and width $M$, using the exact analytic method of 
Beale~\cite{beale1}, which is based on Kaufman's solution of the Ising 
model~\cite{kaufman1}.
We consider both narrow ($M = 3, 4$) and broader strips ($M = 32, 64$).
Both types of strips display novel 
transitions absent for the chain ($M = 1$), though the ones for the broader
strips appear to be closer in nature to those found recently~\cite{qb1,sb1}
for the Ising system on a square lattice ($L^2$). We have no reason to doubt 
that these 
additional transition features of higher order would not survive in 
the thermodynamic limit.
\section{Ising chain and Ising strips}
In the Ising systems we study, the energy of a spin configuration 
$\mathcal{S}=(s_1,s_2,\ldots, s_N)$, with $s_i=\pm 1$, on a 
rectangular lattice with $L$ spins in one direction and $M$ spins
in another ($N=LM$) can be written as
\begin{equation}
\label{eq:ising}
E(\mathcal{S})=-J\sum\limits_{\langle i,j\rangle}s_is_j,
\end{equation}
where $\langle i,j\rangle$ indicates that only interactions of 
nearest-neighbor spin pairs $s_i$ and $s_j$ are considered. 
For $J>0$, ferromagnetic coupling 
energetically favors spins in the same state, whereas for antiferromagnetic 
coupling ($J<0$), configurations of alternating spins are energetically 
preferred. For the discussion of the microcanonical results it is 
already useful to  note that for odd 
choices of $L$ and/or $M$
the total numbers of states with negative and 
positive energies, respectively, are not identical. Thus, the 
athermal energy distribution $g(E)$ (``density'' of states\footnote{In fact, 
since the problem is discrete, one should call this quantity the ``number of 
states,'' but density of states is a more commonly used term even in this 
context.}) cannot be symmetric.

In this study, we re-visit the one-dimensional Ising chain with $L$ spins 
($M=1$) and periodic boundary conditions (rings). For the two-dimensional 
Ising 
strips, we investigate the cases $M=3$ and 
$M=4$, respectively, with periodic boundary conditions in both 
directions (tori).\footnote{We do not consider $M=2$, because periodic 
boundary conditions are somewhat ambiguous in this case.} Eventually, this 
analysis is extended to broader strips with $M=32$ and $M=64$.

The exact solutions for the quantities needed for the statistical analysis 
are known, at least in principle. Revisiting the 1D case is useful for 
comparison. 
For the 2D Ising systems, the calculation of microcanonical 
quantities like the density of states requires algorithmic 
procedures~\cite{beale1} and for large systems it is common to use stochastic 
methods like Monte Carlo sampling to obtain estimates for the 
density of states~\cite{landau1}. Here, we consider only finite systems and 
all analyses are exact. In the 
following, we first investigate the 1D Ising case from the perspective of 
microcanonical statistical analysis in search of transition signals and then 
study transition features of the 2D strips.
\section{Canonical Statistical Analysis of the One-Dimensional Ising Chain}
\begin{figure}
\centerline{\includegraphics[width=8.8cm]{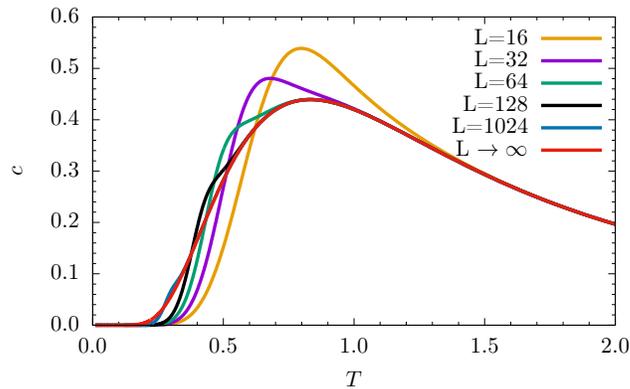}}
\caption{\label{fig:specheat1D} Exact specific heat curves $c(T)$ for the 
one-dimensional Ising chain for 
various chain lengths $L$.}
\end{figure}

For $M=1$, the energy function of the Ising model~(\ref{eq:ising}) represents 
the 1D Ising chain with $L$ spins:
\begin{equation}
E(\mathcal{S})=-J\sum\limits_{l=1}^L s_ls_{l+1},
\end{equation}
where $s_{L+1}=s_1$ to satisfy periodic boundary conditions. The 
canonical partition function, which is the basis for the conventional 
canonical statistical analysis, is given by 
\begin{equation}
Z=\sum\limits_{\{\mathcal{S}\}}e^{-\beta_\mathrm{th} E(\mathcal{S})}
\label{eq:part_sum}
\end{equation}
where $\beta_\mathrm{th}=1/k_\mathrm{B}T$ is the inverse thermal energy. The 
sum is over 
the complete space of spin configurations. Using the eigenvalues of the 
transfer matrix
\begin{equation}
\lambda_+=2\cosh(\beta_\mathrm{th} J),\quad \lambda_-=2\sinh(\beta_\mathrm{th} 
J)
\label{eq:part_sumB}
\end{equation}
the solution can be written as
\begin{equation}
Z=\lambda_+^L+\lambda_-^L.
\end{equation}
Since $\lambda_+>\lambda_-$, the contributions of $\lambda_-$ to 
thermodynamic quantities vanish in the thermodynamic limit $L\to\infty$. This 
approximation is typically used to show that there cannot be a phase 
transition in the 1D case~\cite{wezel1}. However, for our following 
discussion, which 
includes finite-size properties, it is necessary to keep working with the 
full solution. The internal energy of the system is given by
\begin{equation}
\langle E\rangle=-\frac{\partial}{\partial\beta_\mathrm{th}}\ln Z.
\end{equation}
From it, a somewhat lengthy expression for the specific heat (heat capacity 
per spin)
\begin{equation}
c=\frac{1}{L}\frac{\partial \langle E\rangle}{\partial T}.
\end{equation}
can easily be obtained after some algebra. In the thermodynamic limit,
\begin{equation}
\lim_{L\to\infty}c = 
k_\mathrm{B}\beta_\mathrm{th}^2J^2\mathrm{sech}^2(\beta_\mathrm{th} J).
\end{equation}
Curves 
of $c(T)$ are shown in Fig.~\ref{fig:specheat1D} for various finite chain 
lengths $L$ and in the thermodynamic limit $L\to\infty$. Already for $L=1024$, 
the curve is virtually indistinguishable from the well-known result in the 
thermodynamic limit. 

One interesting aspect is the way the peak for $L=16$ disappears, because it 
does not actually seem to converge to the peak location apparent for $L\to 
\infty$. Rather, with increasing chain length, it drifts to lower 
temperatures and the peak height decreases. For $L=64$, only a 
prominent ``shoulder'' remains (and the peak at about $T=0.8$, which 
survives for $L\to\infty$, has formed). 
Increasing $L$ further, the 
``shoulder'' becomes weaker and keeps shifting toward lower temperatures. For 
$L\to\infty$ it is completely gone. The 
height of the remaining peak is finite in this limit. As expected, 
the 1D Ising chain does not 
experience a thermodynamic phase transition. Thus, as a general 
conclusion, a peak in a specific-heat curve does 
\emph{not necessarily} signal a significant qualitative change of the 
thermodynamic macrostate of the system at the temperature associated with the 
specific heat peak.

Now the problem of this argumentation is, however, that for systems, which 
are necessarily of finite size (like biologically relevant 
macromolecules or nanoscale devices), peaks or 
``shoulders'' in specific heat curves and other response quantities 
are the only available signals and often indeed considered important 
indicators of structural transitions in such systems. In heterogeneous systems 
like proteins, finite-size scaling analysis is not applicable and thus it is 
not possible to test if a non-analyticity might develop in the hypothetical 
thermodynamic limit. On the other hand, the typically highly cooperative 
qualitative changes in these systems can be substantial, and it is more than 
just tempting to consider these processes analogs of phase transitions in 
finite systems\footnote{It has been common to call representative ensembles of 
macrostates in finite systems ``pseudophases'' and the crossovers between 
these ``pseudophase transitions,'' but this is a rather unsatisfying way of 
dealing with this problem.}. In consequence, conventional canonical 
statistical analysis cannot resolve this dilemma. The recently developed 
generalized microcanonical inflection-point analysis method~\cite{qb1} was 
introduced to offer an alternative, consistent, and more systematic approach 
to identify and even classify transitions in systems of any size. After a 
brief introduction, we apply it to the 1D Ising chain, which will allow us to 
decide whether or not the peaks and shoulders in the specific heat curves in 
Fig.~\ref{fig:specheat1D} might indicate a hidden transition that simply does 
not develop into a phase transition or whether these canonical features are 
indeed 
irrelevant in the context of macroscopic cooperative behavior. 
\section{Generalized microcanonical inflection point analysis}
The microcanonical entropy version of Boltzmann's formula,
\begin{equation}
S(E)=k_\mathrm{B}\ln g(E),
\end{equation}
has long been used as an alternative basis
for identifying first-order transition signals, which, for finite 
systems, show a distinctive ``convex intruder'' in the otherwise 
strictly concave monotonic behavior of this quantity. In the thermodynamic 
limit, the convex region disappears and becomes linear~\cite{gross1}. 
Considering inflection points in the first 
derivative of $S(E)$, i.e., in the microcanonical inverse temperature 
\begin{equation}
\beta(E) = \frac{dS(E)}{dE},
\label{eq:beta}
\end{equation}
as indicators enabled the extension to second-order transitions~\cite{sslb1}. 
The recently introduced generalized microcanonical inflection-point analysis 
method makes use of the principle of least 
sensitivity~\cite{stevenson1,stevenson2}. Least-sensitive inflection 
points in all derivatives of $S(E)$ are analyzed systematically to identify 
and 
classify transitions of any order~\cite{qb1}, similar to Ehrenfest's 
classification approach, which is based on thermodynamic 
potentials~\cite{ehrenfest1}. However, the great advantage of the novel 
microcanonical method is that it can be employed for systems of any size and 
does not hinge on the thermodynamic limit.

In fact, the consequent application of the method leads to the introduction 
of two types of transitions: \emph{Independent} transitions, which are not 
associated with any other transition signals, and \emph{dependent} 
transitions that can only coexist with a corresponding independent 
transition. A dependent transition, if it exists, can only occur at a 
higher energy (or temperature) and has always a higher order 
than the independent transition it is associated with. Consequently, there's 
no 
first-order dependent transition. These dependent transitions can be 
considered valuable precursor signals in the  
disordered phase of imminent ordering upon lowering system energy (or 
temperature).

Formally, a phase transition is defined in the generalized microcanonical 
inflection-point analysis method by a corresponding least-sensitive 
inflection point in $S(E)$ or any of its derivatives at the 
transition energy $E_\mathrm{tr}$. For practical purposes, 
it is useful to identify the extremal point in the next-higher derivative. 
Consequently, for \emph{independent transitions} of odd order $(2k-1)$ ($k$ 
positive integer), the valley of the $(2k-1)$th derivative has 
a positive-valued minimum at the transition energy $E_\mathrm{tr}$,
\begin{equation}
\left. \frac{d^{(2k-1)}S(E)}{dE^{(2k-1)}} 
\right|_{E=E_{\mathrm{tr}}}>0, 
\label{eq:odd_order_trans}
\end{equation}
and for even order $2k$ the peak of the $(2k)$th derivative at 
$E_\mathrm{tr}$ is negative-valued:
\begin{equation}
\left. \frac{d^{2k}S(E)}{dE^{2k}} \right|_{E=E_{\mathrm{tr}}}<0. 
\label{eq:even_order_trans}
\end{equation}
The lowest-possible order of a \emph{dependent transition} is 2. In 
general, dependent transitions of even order $2k$ are characterized by
\begin{equation}    
\left. \frac{d^{2k}S(E)}{dE^{2k}} \right|_{E=E_\mathrm{tr}^\mathrm{dep}}>0, 
\label{eq:even_order_dep_trans}
\end{equation}
whereas for odd order $(2k+1)$:
\begin{equation}
\left. \frac{d^{(2k+1)}S(E)}{dE^{(2k+1)}}
\right|_{E=E_\mathrm{tr}^\mathrm{dep}}<0. 
\label{eq:odd_order_dep_trans}
\end{equation}
must be satisfied. For brevity, we introduce in addition to 
$\beta(E)$ given in Eq.~(\ref{eq:beta}) the following symbols for
higher-order derivatives: $\gamma(E)=d^2S(E)/dE^2$, $\delta(E)=d^3S(E)/dE^3$,
$\epsilon(E)=d^4S(E)/dE^4$, etc.

\begin{figure}
\centerline{\includegraphics[width=8.8cm]{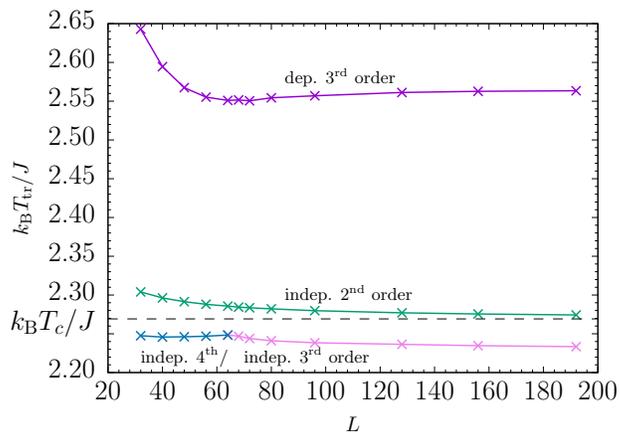}}
\caption{\label{fig:Ttr2D} The recent microcanonical inflection-point 
analysis for the full 2D Ising model~\cite{qb1,sb1} correctly identified the 
second-order transition, which becomes critical at 
$k_\mathrm{B}T_c/J=2/\ln(1+\sqrt{2})$ in the thermodynamic limit, but two 
additional higher-order signals were found as well (lines are guides to the 
eye).}
\end{figure}

Remarkably, for the 2D Ising model with $L^2$ spins, not only the 
expected ferromagnetic-paramagnetic transition was recovered and correctly 
classified as a second-order transition with this method. Two additional 
transitions, a dependent transition above and an independent transition below 
the critical temperature were discovered~\cite{qb1,sb1}. For sufficiently 
large 
systems, these transitions are of third order and the extrapolation of the 
exact data calculated for finite systems with up to $N=192^2$ spins suggests 
that 
they would survive in the thermodynamic limit (see 
Fig.~\ref{fig:Ttr2D}).
\subsection{Statistical analysis for the 1D Ising chain}
\begin{figure}
\centerline{\includegraphics[width=16cm]{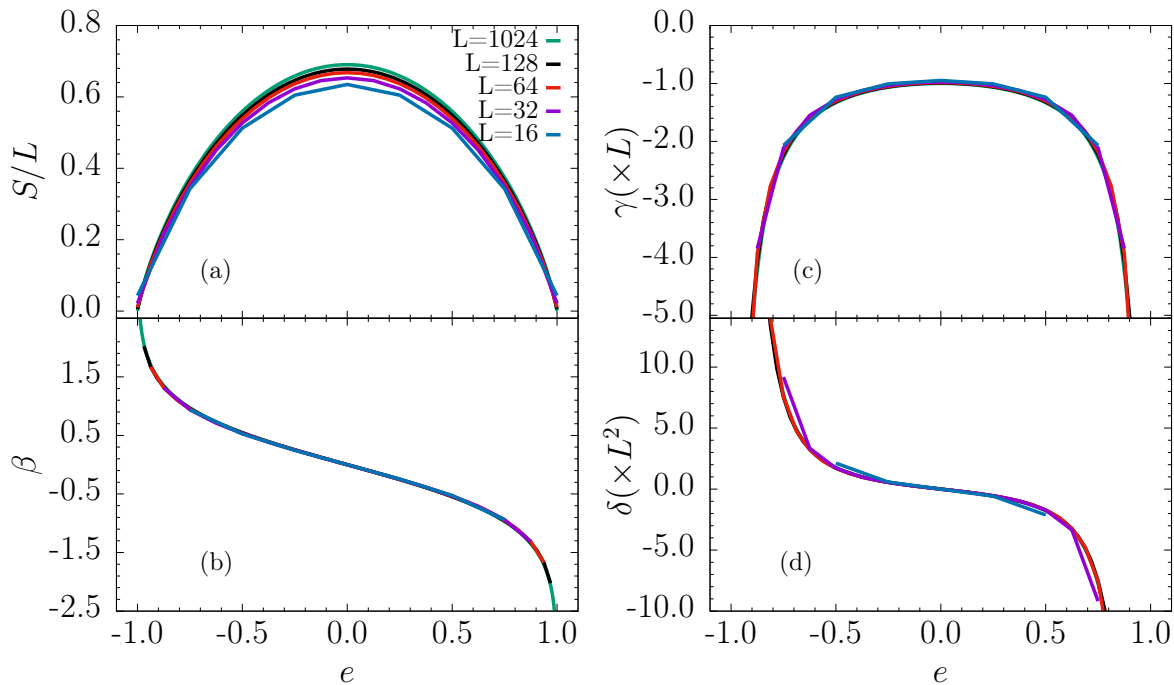}}
\caption{\label{fig:micro1D} Microcanonical curves of 
the 1D Ising chain for different chain lengths $L$ as a function of 
$e=E_b/L$ (lines are guides to the eye.)}
\end{figure}

The inflection-point analysis for the 1D model is straightforward and all 
results for any spin chain length are easily found analytically. 
For simplicity, 
we consider an even number of spins $L$ and periodic boundary conditions. It 
is convenient to introduce the number $b$ of ``broken'' bonds. 
A single broken bond represents an energy increase by $2J$. Now,
a spin flip affects a pair of bonds and only if both are 
broken after the flip, the energy increases. Thus, $b$ must be an even 
integer. The total energy of a spin configuration
is then given by
\begin{equation}
E_b=(-L+2b)J, \quad b=0,2,4,\ldots,L. 
\end{equation}
Simple combinatorics yields the number of 
configurations $g_b$ for $b$ broken bonds. We exploit the route
of the canonical partition function~(\ref{eq:part_sum}), since it can be 
expressed as
a sum over the number of states $g_E$ with given energy $E$:
\begin{equation}
Z=\sum\limits_{E} g_E\, e^{-\beta_\mathrm{th} E} = \sum\limits_{b} g_b\, 
e^{-\beta_\mathrm{th} E_b}.
\end{equation}
Making use of the solution~(\ref{eq:part_sumB}) and the binomial expansion 
yields
\begin{equation}
Z=2\sum\limits_{b=0,2,4,\ldots}^L 
\left(\begin{array}{c}
L\\ b
\end{array}\right)\, e^{-\beta_\mathrm{th}E_b}
\end{equation}
and thus
\begin{equation}
g_b=2\left(\begin{array}{c}
L\\ b
\end{array}\right),
\end{equation}
as expected. With this, we can define the microcanonical entropy
\begin{equation}
S_b=k_\mathrm{B}\ln\, g_b.
\end{equation}
Curves of $S_b$ are shown in 
Fig.~\ref{fig:micro1D}(a) for several chain lengths.

As analogs of the continuous derivatives, we employ the following discrete, 
symmetric differences:
\begin{eqnarray}
\beta_{E_b}&=&\frac{S_{b+2}-S_{b-2}}{2\Delta E_b},\\
\gamma_{E_b}&=&\frac{S_{b+2}-2S_b+S_{b-2}}{(\Delta E_b)^2},\\
\delta_{E_b}&=&\frac{S_{b+4}-2S_{b+2}-2S_{b-2}-S_{b-4}}{2(\Delta 
E_b)^3},\\
\end{eqnarray}
where $\Delta E_b=4J$. For the subsequent analysis, we set the irrelevant 
constants to unity, $J\equiv 1$ and $k_\mathrm{B}\equiv 1$.

If there was a first-order transition, $S_b$ should have a 
least-sensitive inflection point indicated by a positive-valued minimum in 
\begin{equation}
\beta_{E_b}=\frac{1}{8}\ln\frac{(L-b+2)(L-b+1)(L-b)(L-b-1)}{(b+2)(b+1)b(b-1)},
\end{equation}
which obviously does not exist for the 1D Ising chain. Introducing the 
parameter $x_b=b/L$, Taylor expansion yields for $0<x_b<1$ (i.e., 
explicitly excluding the edges $x_b=0$ and $x_b=1$)
\begin{equation}
\beta_{E_b}=\frac{1}{2}\ln\frac{1-x_b}{x_b}+\frac{1}{4}
\frac{2x_b-1}{x_b(1-x_b)}\frac{1}{L}+\mathcal{O}(1/L^2).
\end{equation}
Thus, in the thermodynamic limit $L\to\infty$, $\beta_{E_b}$ converges to 
$(1/2)\ln[(1-x_b)/x_b]$ for any given value $x_b\in(0,1)$.
Figure~\ref{fig:micro1D}(b) shows the 
$\beta_{E_b}$ curves for various chain lengths $L$. The quick 
convergence of the 
curves to the result in the thermodynamic limit is already apparent for small 
chain lengths. In the typically only considered negative-energy region, 
associated with positive temperatures, $\beta_{E_b}$ does not exhibit 
inflection points at any chain length. The same applies in the generally 
ignored positive-energy (negative-temperature) space. However, since 
$\beta_{E_b}$ is convex for $E_b<0$ and concave for $E_b>0$, it possesses a 
least-sensitive inflection point at $E_b=0$. This is confirmed by the 
associated peak in 
\begin{eqnarray}
\gamma_{E_b}&=&\frac{1}{16}
\ln\frac{(L-b)(L-b-1)b(b-1)}{(L-b+2)(L-b+1)(b+2)(b+1)}\\
&=&-\frac{1}{4}\frac{1}{x_b(1-x_b)}\frac{1}{L}+\mathcal{O}(1/L^2). 
\end{eqnarray}
According to our classification scheme, this signals an independent, but 
thermodynamically irrelevant, second-order transition in the 1D model at 
$\beta = 0$ that does not disappear in the thermodynamic limit.  
Note that, despite formally being of second order in our scheme, 
this is not a critical transition; the appropriately 
scaled derivative $\gamma\times L$ does not vanish in the 
thermodynamic 
limit [see Fig.~\ref{fig:micro1D}(c)]. Although this example is not 
particularly interesting, it shows that discontinuous, critical transitions 
represent only a subset of second-order transitions in this scheme.

Eventually, we find
\begin{eqnarray}
\delta_{E_b}&=&\frac{1}{128}\ln\frac{(L-b+4)(L-b+3)(L-b-2)(L-b-3)}
{(L-b+2)(L-b+1)(L-b)(L-b-1)} \nonumber \\ 
&&\hspace*{12mm}\times \frac{(b+2)(b+1)b(b-1)}{(b+4)(b+3)(b-2)(b-3)}, \\
&=&-\frac{1}{8}\frac{2x_b-1}{x_b^2(1-x_b)^2}\frac{1}{L^2}+\mathcal{O}(1/L^3)
\quad \forall x_b\in(0,1)
\end{eqnarray} 
and, as shown in Fig.~\ref{fig:micro1D}(d), the scaled $\delta\times L^2$ 
curves do not 
reveal any features of higher-order transitions.

To conclude, microcanonical inflection-point analysis does not uncover any 
transition features among the equilibrium states of the 1D Ising model. This 
confirms the expected result from conventional analysis, although response 
quantities like the specific heat show pronounced thermal activity, which in 
fields like the polymer sciences is often considered a sign of significant 
cooperative behavior. This can lead to 
confusion or an ambiguity of interpretations for finite systems, which the 
microcanonical analysis method used here does not allow.
\subsection{Transition properties of narrow Ising strips}
In order to investigate the onset of the paramagnetic-ferromagnetic phase 
transition in the 2D Ising model, we extend our consideration from the 1D 
Ising chain (ring) to 2D 
strips (tori with finite tube diameter). We initially choose to study the 
$L\times 3$ and 
$L\times 4$ systems (with $L$ even as before) with periodic boundary 
conditions in 
both dimensions. This will allow us to discuss potential differences in the 
phase behavior for systems that are finite in the second dimension and 
(non)symmetric under periodic boundary conditions. 
Later on, we will study the impact of $M$ on the transition signals by 
investigating broader strips.

Since the focus will be on the microcanonical analysis and interpretation, we 
use the algorithmic solution by Beale for the exact number of states 
$g_E$~\cite{beale1} to 
calculate the microcanonical entropy and its derivatives needed for our 
analysis. This method makes use of the low-temperature expansion for 
the partition function of the $L\times M$ Ising model with periodic boundary 
conditions, which was first calculated by Kaufman~\cite{kaufman1}.

\begin{figure}
\centerline{\includegraphics[width=16cm]{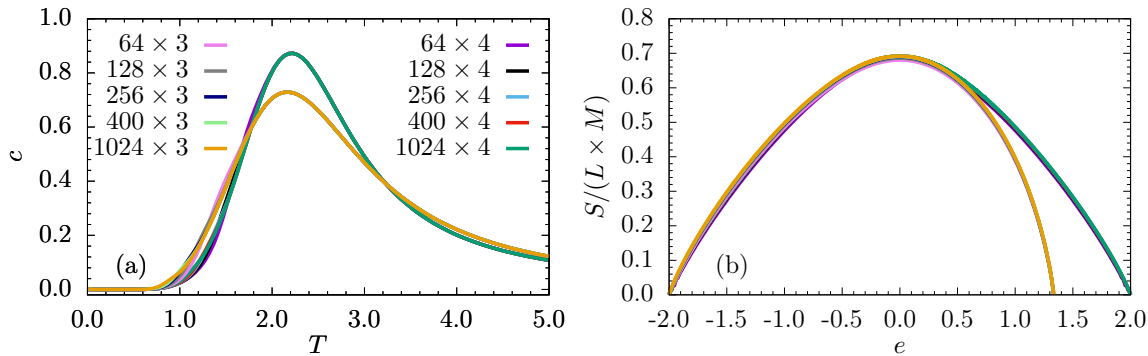}}
\caption{\label{fig:candS3_4} (a) Specific heat curves $c(T)$ and (b) 
microcanonical entropies $S(e)$ with $e=E/LM$ for $L\times M$ 
Ising strips with $M=3,4$ at various chain lengths $L$. Note that both 
figures contain two sets of curves.}
\end{figure}

For comparison, we also use $g_E$ to determine the canonical 
moments of the energy and from these the specific heat,
\begin{equation}
c=\frac{1}{LMk_\mathrm{B}T^2}\left(\langle E^2\rangle-\langle E\rangle^2 
\right),
\end{equation}
which is shown in Fig.~\ref{fig:candS3_4}(a) for both $M=3$ and $M=4$ and 
various strip lengths $L$. Apparently, the curves for each scenario fall 
almost on top of each other, which, similarly to the 1D case, suggests that 
$c$ does not exhibit singular scaling properties. Broadening
the 1D chain to a narrow strip does not enforce a spontaneous 
breaking that could cause a phase transition (for a recent review on 
spontaneous symmetry breaking, see Ref.~\cite{wezel1}).

It is noteworthy, though, 
that the peak value for the broader strip ($M=4$) is larger than for the 
smaller one, which indicates the development of a transition signal upon 
increasing $M$. We will further elaborate on this in the subsequent 
discussion of broader strips with $M=32,\,64$. For now it is sufficient to 
observe that the sole introduction of a second dimension does not lead to 
a thermodynamic phase transition in the system. 
Conventional canonical 
statistical analysis thus discards this scenario as not 
being thermodynamically interesting. But what does the microcanonical 
analysis yield?

\begin{figure}
\centerline{\includegraphics[width=16cm]{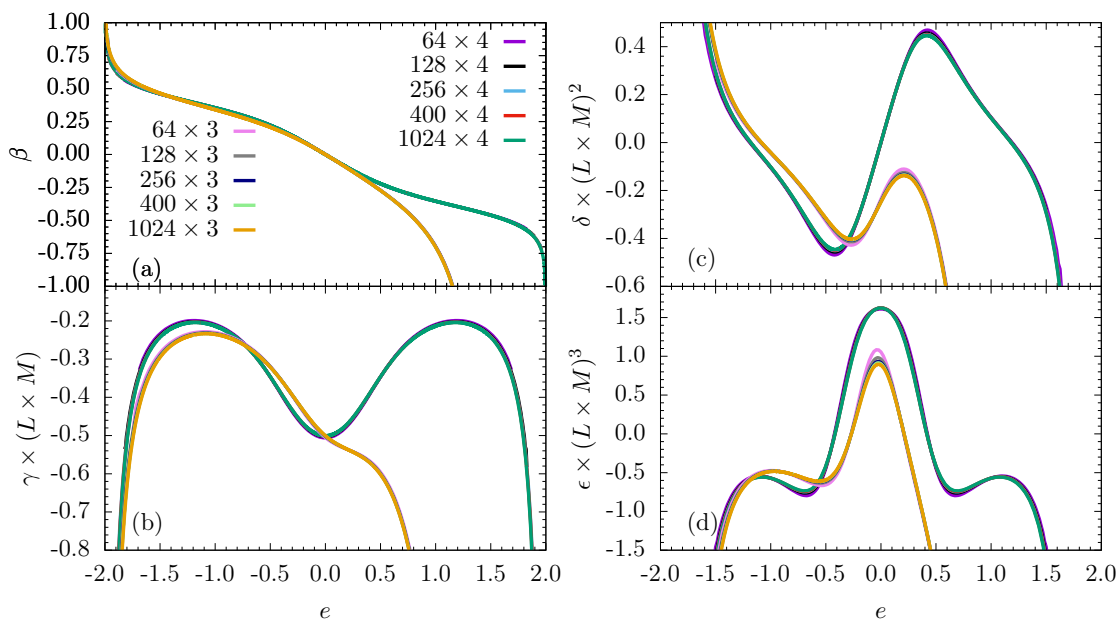}}
\caption{\label{fig:micro3_4} Derivatives of the microcanonical entropies for 
the $L\times 3$ and $L\times 4$ Ising strips.}
\end{figure}

In Fig.~\ref{fig:candS3_4}(b), the microcanonical entropy curves are shown for
both cases. As expected, the curves for $M=3$ are nonsymmetric, because 
fewer positive energy configurations are possible (as they favor 
antiferromagnetic alignments while periodic boundary conditions lead to
frustration). 
Like in the 1D case, the entropies scale with system size $L M$ in the 
reduced energy space $e=E/LM$. The entropy curves do not show significant 
features so let us now focus on the derivatives plotted in 
Figs.~\ref{fig:micro3_4}(a)--(d). In fact, the microcanonical results are very 
intriguing, although they also 
do not show any indication of critical behavior in the thermodynamic limit.

Like in the 1D case, the first derivative $\beta(e)$ is virtually 
scale independent, but in striking contrast to the 1D case, there is no 
least-sensitive inflection point at $e=0$ anymore. Instead, 
in the symmetric case $L\times 4$, it has been replaced by a pair of 
inflection points \emph{within} $e \ne 0$ regions. 
These signals of independent second-order transitions, which 
appear prominently as peaks in the $\gamma$ curves for $L\times 4$ 
[Fig.~\ref{fig:micro3_4}(b)], mark the existence of the boundaries
between the disordered paramagnetic phase (with $e$ around $0$) and 
the familiar ordered ferromagnetic phase (in the $e<0$ region), as well as 
the less familiar antiferromagnetic phase (with $e>0$, corresponding to 
negative microcanonical temperatures). This is the precursor to the known 
behavior 
of the 2D Ising system in the thermodynamic limit (in both dimensions). Hence 
we confirm that this qualitative change in $\beta(e)$ accompanied by the 
creation of a new phase is caused by the sole existence of the second 
dimension.

The nonsymmetric system falls short of developing a second-order transition 
signal in the positive-energy space due to lack of available microstates, 
but the least-sensitive inflection point in $\gamma(e)$ at $e\approx 0.2$ 
suggests a (dependent) third-order transition in addition to the prominent 
second-order signal in the negative-$e$ region. Though we have not studied
systems with odd strip widths $M>3$, there are good reasons to 
believe that 
this asymmetry will fade away as $M$ becomes large.\footnote{We should note 
that,
if we had imposed free boundary conditions instead of periodic ones, 
there would be no asymmetry.}

Both systems possess another independent transition of fourth order (again 
two symmetric signals for $L\times 4$, but one only for $L\times 3$), as 
indicated by the least-sensitive inflection points in $\delta(e)$.

\begin{figure}
\centerline{\includegraphics[width=8.8cm]{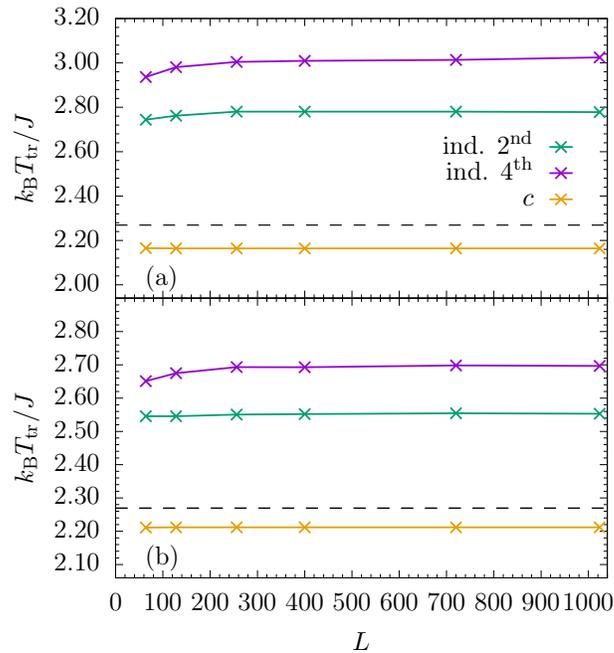}}
\caption{\label{fig:Ttr3_4} Dependence of the microcanonical transition 
temperatures on strip length $L$ for strip 
widths (a) $M=3$ and (b) $M=4$. 
Analyses were done for lengths marked by a cross. For comparison, the 
peak temperatures of the specific heat $c$ are also included in the figures 
for the respective strip parameter values. Lines are guides to the eye 
only. The horizontal dashed lines are located 
at the critical 
temperature of the 2D Ising model in the thermodynamic limit, 
$k_\mathrm{B}T_c/J=2/\ln(1+\sqrt{2})$.}
\end{figure}

Figure~\ref{fig:Ttr3_4} summarizes all information about the transition 
points identified in the canonical and microcanonical analysis of the various 
systems sizes studied. For the discussion of equilibrium behavior only 
non-negative temperatures are relevant. The dashed line is located at the 
critical temperature of the 2D Ising system in the thermodynamic limit and 
serves as a reference. First of all, as already expected, none of the 
transition curves shows any trend to converge to the critical point in the 
thermodynamic limit, but the fact that the signals do not disappear or 
``fade out'' either is remarkable. It tells us that the inherent phase 
structure of the 2D Ising system is already present in the narrow Ising strip 
scenarios, although critical behavior is not achieved even for infinitely long 
strips ($L\to\infty$). The transition temperatures found for $M=4$ are closer 
to the critical point of the 2D Ising model than those identified for $M=3$.
\subsection{Qualitative changes for broader Ising strips}
The results discussed above provoke the question, whether or not there is a 
finite threshold strip width $M$, beyond which the Ising 
strip develops symptoms of critical behavior. In order to keep the ability of 
using the Beale method to obtain exact microcanonical results, we restrict 
our discussion to widths $M=32$ and $M=64$. We also investigate only 
the thermodynamic behavior at positive microcanonical temperatures (i.e., 
negative energies).
\begin{figure}
\centerline{\includegraphics[width=8.8cm]{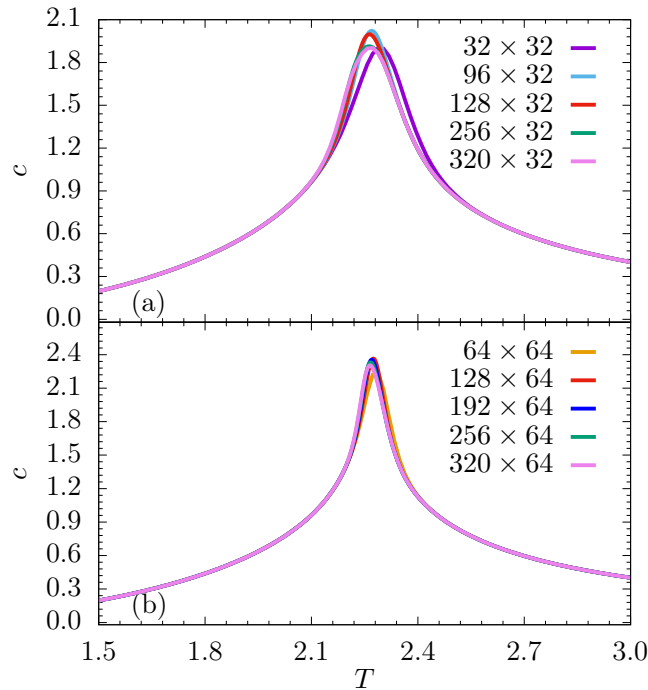}}
\caption{\label{fig:c32_64} Specific heat curves 
$c(T)$ for Ising strips with (a) $M=32$ and (b) $M=64$ at various chain 
lengths $L$.}
\end{figure}

Let us first take a look at the specific-heat curves shown in 
Fig.~\ref{fig:c32_64}. For $M=32$, we observe that initially the peak height 
increases, but for strip lengths $L>96$, it decreases again and
converges to a finite height, similar to the 1D case. If $M=64$, the 
peak of the specific heat is almost independent of $L$, although we still 
notice a slight decrease of the peak height upon increasing $L$ for $L>128$. 
The peak heights are higher and the peak regions narrower for the broader 
strip, supporting the development of the expected singularity at the critical 
temperature in the limits $M\to\infty$ and $L\to\infty$. From the 1D case we 
know that a peak in the specific heat curves does not necessarily mean that 
the system undergoes qualitative macrostate changes. However, the 
microcanonical analysis seems to be more capable of distinguishing 
significant from insignificant fluctuations. In contrast to the 1D case we 
found noticeable transition features for the narrow Ising strips as discussed 
above. This encourages the use of the microcanonical approach for the broader 
strips, too. 

\begin{figure}
\centerline{\includegraphics[width=16cm]{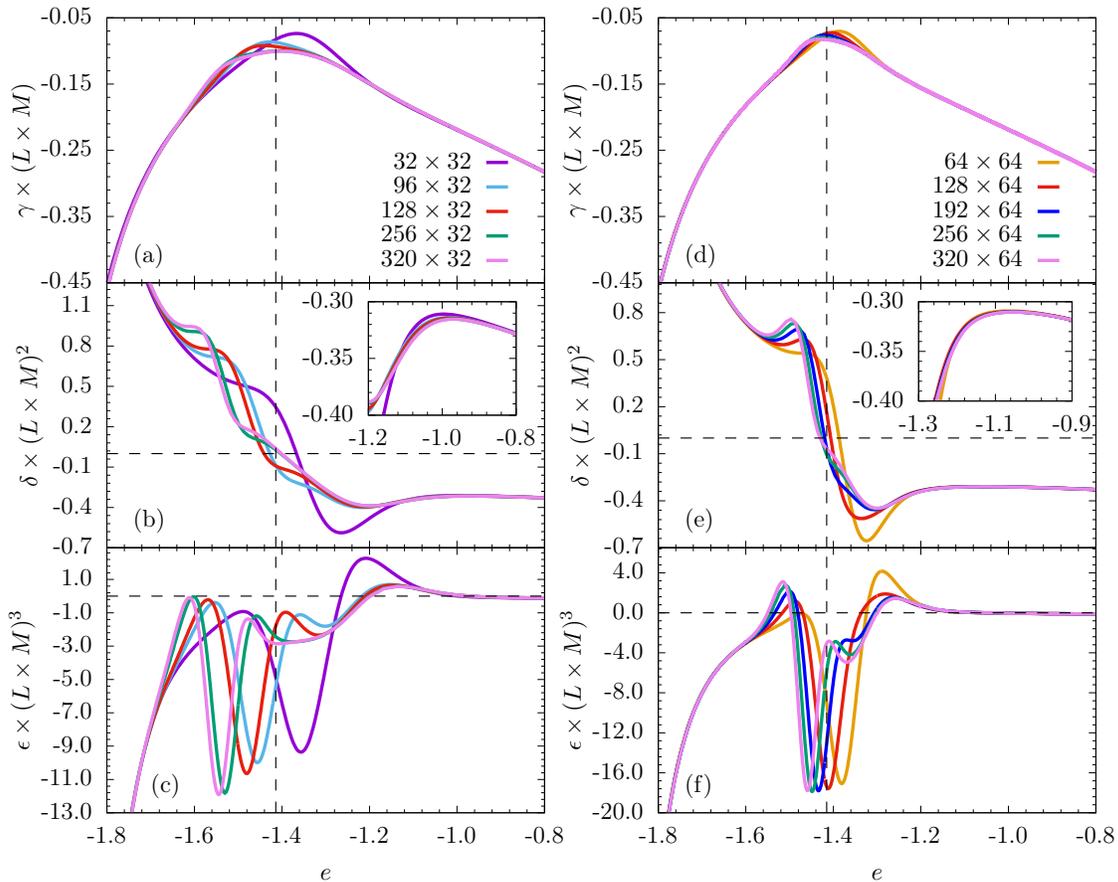}}
\caption{\label{fig:micro32_64} Derivatives of the microcanonical entropies 
for the $L\times 32$ and $L\times 64$ Ising strips. Horizontal dashed lines 
mark zero and vertical dashed lines are located at the transition energy 
associated with the critical transition temperature of the 2D Ising model.}
\end{figure}

\begin{figure}
\centerline{\includegraphics[width=8.8cm]{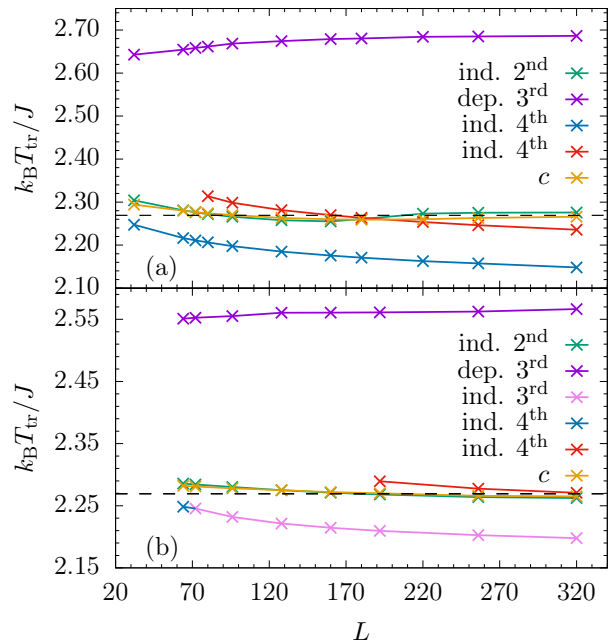}}
\caption{\label{fig:Ttr32_64} Microcanonical transition 
temperatures and peak temperatures of the specific heat curves $c$ 
versus strip length $L$ for (a) $M=32$ and (b) 
$M=64$ 
(lines are guides to the eye). The 
dashed lines mark the critical 
temperature of the 2D Ising model in the thermodynamic limit, 
$k_\mathrm{B}T_c/J=2/\ln(1+\sqrt{2})$.}
\end{figure}

Entropy curves are not shown here, because they do not exhibit any striking 
monotonic feature of interest. Furthermore, the first derivative [or 
$\beta(e)$; not shown either] only exhibits a single obvious least-sensitive 
inflection point 
indicating a second-order transition for all studied systems sizes with 
$M=32$ and $M=64$. The corresponding peaks are clearly present in the 
next-higher derivative $\gamma(e)$ shown in Figs.~\ref{fig:micro32_64}(a) 
and~(d). In this context it is important to notice that the peak height 
drops, which effectively means that the inflection point in $\beta(e)$ 
becomes \emph{more sensitive the larger the system size}. This leads to the 
conclusion that whereas the second-order signal is prominent and does not 
show any tendency of disappearing in the thermodynamic limit, the transition 
is not ``thermodynamic'' in the traditional sense. For this to be the case, 
the peak in $\gamma(e)\times LM$ must converge to zero, which essentially 
means that the inflection point in $\beta(e)$ becomes a saddle point and thus 
\emph{minimally sensitive}. The transition signal we observe, however, is 
remarkable as the peak locations converge toward the 2D Ising transition 
point in energy space (vertical dashed line in Fig.~\ref{fig:micro32_64} 
associated with 
the critical point in temperature space). Figure~\ref{fig:Ttr32_64} shows 
the microcanonical transition temperatures versus the strip 
lengths for both $M=32$ and $M=64$. The second-order signals very quickly 
converge to the critical temperature (dashed line) for larger systems. Thus, 
the second-order transition found for the broader Ising strips is the 
analog of the critical transition in the 2D Ising model. Note that such 
convergence was not observed for the narrower strips 
[cf.~Fig.~\ref{fig:Ttr3_4}].

In addition to the prominent second-order transition a number of higher-order 
transition signals can be identified. The $\gamma$ curves in 
Figs.~\ref{fig:micro32_64}(a) and~(d) already hint at two other significant 
transitions; one in the upper-critical and the other in the subcritical 
regime. 

Employing our classification scheme, the lowest-order transition at 
energies above the critical energy (or, equivalently, temperature above the 
critical temperature) is of third order and its existence depends on the 
second-order transition described above. As Figs.~\ref{fig:Ttr32_64}(a) 
and~(b) show, the transition point does not show any convergence toward the 
critical point and, like in the 2D Ising case~\cite{qb1,sb1}, we conclude 
that 
this is a separate transition in the paramagnetic phase. Since it is dependent 
on the second-order transition, one may want to interpret it a precursor of 
the latter. As the insets in the 
respective Figs.~\ref{fig:micro32_64}(b) 
and~(e) of the third entropy derivative $\delta(e)$ show, the peak heights 
associated with this transition signal are very robust and do not change much 
upon increasing the strip length.

The other noteworthy transition develops in the subcritical regime. It is 
signaled by the
lowest-energy peaks in Fig.\ref{fig:micro32_64}(c) near $e\approx -1.6$ for 
$M=32$ and the developing local minimum in Fig.~\ref{fig:micro32_64}(e) at 
about $e\approx -1.55$ for $M=64$. It is an independent transition of fourth 
order for all strip lengths studied and width $M=32$, and turns into a 
third-order 
independent transition for $M=64$ if $L\ge 72$. As can be seen in 
Figs.~\ref{fig:Ttr32_64}(a) and (b), this transition also seems to converge 
toward a 
transition point away from the critical point. This 
transition line was observed in the 2D Ising case as 
well~\cite{qb1,sb1}, see Fig.~\ref{fig:Ttr2D}. In fact,
as exemplary tests we performed show, strips with $M\ge 64$ behave 
very similar to the full 2D Ising system; 
the second-order transition (which becomes critical in the thermodynamic 
limit) and the two additional third-order transitions consolidate.

These two transitions seem to bracket what one might want to call the 
``critical atmosphere'' around the critical point. There is an additional 
independent fourth-order transitions in the vicinity of the critical point 
which we included in Figs.~\ref{fig:Ttr32_64}(a) and (b). For $M=32$ we 
clearly see it crossing the critical transition, which may indicate a 
qualitative change in the system that is unrelated to the 
ferromagnetic-paramagnetic transition.

Higher-than-fourth-order transitions were not included in our analysis, 
because the relevance of a transition signal diminishes with increasing 
order. Nonetheless, an in-depth discussion of the individual
transition behavior and characteristics of Ising systems may require the 
consideration of additional precursor signals.

Finally, we would like to remark that the peaks observed in the specific heat 
curves for the broader Ising strips in Fig.~\ref{fig:c32_64} indeed indicate 
the onset of the critical transition. As the 
$c$ line in Fig.~\ref{fig:Ttr32_64} shows, the 
peak temperatures converge to the critical 
temperature known from the 
2D Ising system. This was neither the case for the 1D system nor for the 
narrow strips with $M=3$ and $M=4$, which suggests that the 
qualitative cooperative behavior of spins in Ising strips changes between 
$M=4$ and $M=32$. 
\section{Conclusion}
By means of exact statistical analysis, we identified and classified 
transition features in ferromagnetic one-dimensional Ising chains and 
two-dimensional Ising strips for various finite sizes. For this purpose, we 
employed the algorithmic approach proposed by Beale~\cite{beale1} for the 
exact evaluation of the number of states. The logarithm of this quantity can 
be interpreted as the microcanonical (energy-dependent) entropy. We then used 
the generalized microcanonical least-sensitive inflection-point analysis 
method~\cite{qb1} to systematically identify and classify transition signals 
in the entropy and its derivatives. We contrasted these results with features 
represented by extremal energetic fluctuations, i.e., maxima in corresponding 
(canonical) specific-heat curves, which are often used as indicators of 
transitions in systems of finite size (most prominently in thermodynamic 
studies of biological processes such as the folding of proteins, for which the
thermodynamic limit is nonsensical due to the intrinsic disorder in their 
primary amino-acid sequence).

As expected, the microcanonical analysis of the one-dimensional Ising chain 
did not reveal any transition signals. Neither 
could we locate least-sensitive inflection points for finite chain lengths, nor 
did the extrapolation toward larger systems hint at potential transitions for 
the infinitely long chain. This is consistent with the historic findings for 
this system. It is worth noting that the specific heat curves 
in fact \emph{do exhibit a peak}, which converges to a finite value in the 
thermodynamic limit. However, this extremal energetic fluctuation cannot be 
associated with any transition feature, even for finite systems, as there is 
no support for it by microcanonical analysis. Thus, we conclude that 
considering only extremal fluctuations in canonical response quantities as 
indicators for transitions may be 
misleading in the analysis of qualitative macroscopic
changes in finite systems upon changing external state variables such as the 
canonical (heat-bath) temperature. 

We then turned our attention to effectively two-dimensional Ising strips. For 
different constant strip widths $M=3,4,32,64$ we systematically extended the 
strip lengths $L$ to find trends for transition signals. Like in the 1D 
case, we exclusively used the exact Beale solutions for the density of states 
to avoid numerical errors (the use of Monte Carlo simulations and finite-size 
scaling analysis to consolidate our results is a future project). In all 
cases studied, the maxima of the specific heat curves do not show any trend 
to grow beyond finite limiting values. In fact, the narrower strips ($M=3,4$) 
behave like a one-dimensional Ising chain. For the broader strips 
($M=32,64$), the peak region is significantly narrower, suggesting a 
qualitative change in the crossover regime between paramagnetic and 
ferromagnetic states. As expected, the specific-heat curves are not very 
helpful in revealing more specific information.

However, microcanonical inflection-point analysis enables a more diverse 
discussion. For all strip widths, and in contrast to the 1D case, we find a 
second-order transition, which we expect to be the analog of the critical 
transition in the full 2D case. For the narrower strips, the transition 
temperature does not show any tendency to converge toward the critical 
temperature, but settles at a larger value. Remarkably, the results for the 
broader Ising strips suggest that the observed second-order transition is 
closely related to the critical transition, despite the finite widths of the 
strips. In these cases, the second-order transition temperatures do converge 
to the critical value as the strip length is increased. For the broader 
strips, the microcanonically determined second-order transition temperatures 
correspond very well to the peak locations of the specific heat curves.

Like for the 2D Ising model on the square lattice~\cite{qb1,sb1}, we 
identify a 
number 
of additional transitions of higher order. Most striking are the 
dependent third-order transitions in the paramagnetic phase for $M=32,64$, 
which do not show indications of disappearing in the thermodynamic 
limit, 
although they do not seem to be linked to any catastrophic singularity in the 
thermodynamic limit either, which is probably the reason why to the best of 
our 
knowledge these signals have not been reported until recently~\cite{qb1,sb1}. 
Since this transition is inevitably linked to the second-order transition, we 
consider it a precursor of the latter in the paramagnetic phase. 

The lowest-temperature transitions we find occur in the ferromagnetic phase 
of the $M=32$ and $M=64$ systems and are independent of the critical 
transition. There is no tendency of the transition line to merge with the 
critical (or second-order) transition in the thermodynamic limit either. For 
$M=32$ this transition is of fourth order, but rises to third order for 
sufficiently long $M=64$ strips. This transition does also exist for the 
Ising system on the square lattice~\cite{qb1,sb1}.

The investigation of these transitions for larger systems requires Monte 
Carlo simulations and is left to future work. This will also allow the 
characterization of the additional (non-critical) transitions by 
systematically analyzing structural properties of spin configurations.
\section*{Acknowledgments}
We would like to thank Kevin Bassler for helpful discussions.
\section*{References}
\end{document}